\pdfoutput=1
\documentclass[11pt,reqno,preprint]{article}
\usepackage{jheppub,epsfig,amssymb,amsmath,mathrsfs,hyperref}
\usepackage{multirow,bbm}
\usepackage[makeroom]{cancel}
\usepackage{ytableau}
\usepackage{xcolor}
\usepackage{tikz}
\usepackage{comment}
\usepackage{float}

\definecolor{almond}{rgb}{0.94, 0.87, 0.8}
\definecolor{crimson}{rgb}{0.86, 0.08, 0.24}
\definecolor{mintcream}{rgb}{0.96, 1.0, 0.98}
\definecolor{champagne}{rgb}{0.97, 0.91, 0.81}
\definecolor{bubblegum}{rgb}{0.99, 0.76, 0.8}
\definecolor{airforceblue}{rgb}{0.36, 0.54, 0.66}
    	\definecolor{almond}{rgb}{0.94, 0.87, 0.8}
    	\definecolor{antiquefuchsia}{rgb}{0.57, 0.36, 0.51}
    	\definecolor{antiquewhite}{rgb}{0.98, 0.92, 0.84}
\newcommand{\BoxF}{ {\ytableausetup{boxsize=0.5em}\ydiagram[*(antiquefuchsia)]{1}  \ytableausetup{boxsize=1em} } }
\newcommand{\cut}{\operatorname{cut}}
\def\be{\begin{equation}}
\def\ee{\end{equation}}
\def\ba{\begin{eqnarray}}
\def\ea{\end{eqnarray}}
\newcommand{\bea}{\begin{eqnarray}}
\newcommand{\eea}{\end{eqnarray}}

\def\Li{\textrm{Li}}
\def\l{\langle}

\newcommand{\tx}{\tilde{x}}
\newcommand{\bfx}{\mathbf{x}}
\newcommand{\ev}[1]{\Big\langle #1 \Big\rangle}
\DeclareMathOperator{\bfSh}{\hat{\bf S}}
\newcommand\sfC{\mathsf{C}}

\newcommand{\cP}{\begin{cal}P\end{cal}}

 \newcommand{\Rem}[1]{\operatorname{Rem}(#1)}
 \newcommand{\Add}[1]{\operatorname{Add}(#1)}
\newcommand{\zb}{\bar{z}}

\newcommand{\az}[1]{{\color{azure(colorwheel)}#1}}
\newcommand{\gr}[1]{{\color{green(html/cssgreen)}#1}}
\newcommand{\cri}[1]{{\color{crimson}#1}}
\definecolor{bleudefrance}{rgb}{0.19, 0.55, 0.91}

\newcommand{\bfS}{{\bf S}}
\newcommand{\hbfS}{\hat{\bfS}}

\preprint{BONN-TH-2026-19}

\title{Antipodal Self-Duality for Generalized Fishnets from Integrability}

\author{Lance~J.~Dixon$^{1}$,}
\author{Claude Duhr$^{2}$,}
\author{Francesca~C.~Fernandes$^{3,4}$,}
\author{and Victor Mishnyakov$^{5}$}

\affiliation{$^1$ SLAC National Accelerator Laboratory, Stanford University, Stanford, CA 94309, USA}
\affiliation{$^2$ Bethe Center for Theoretical Physics and Cluster of Excellence ``Color meets Flavor'', Universit\"{a}t Bonn, D-53115 Bonn, Germany}
\affiliation{$^3$ Physics Department, Stanford University, Stanford, CA 94309, USA}
\affiliation{$^4$ Rudolf Peierls Centre for Theoretical Physics, University of Oxford, Clarendon Laboratory, Parks
Road, Oxford OX1 3PU, United Kingdom}
\affiliation{$^5$ 
Moscow Institute of Physics and Technology, 141700, Dolgoprudny, Russia}
\emailAdd{lance@slac.stanford.edu}
\emailAdd{cduhr@uni-bonn.de}
\emailAdd{ffernand@caltech.edu}
\emailAdd{mishnyakovvv@gmail.com}

\abstract{Antipodal self-duality (ASD) is a remarkable property that relates an amplitude to itself, after performing a suitable kinematic map and reversing the letters of the symbol (more precisely, acting with the antipode of the Hopf algebra of the de Rham analogs of multiple polylogarithms).
ASD has been seen in only one process in planar $\mathcal{N}=4$ super-Yang-Mills theory, and its physical origin is unknown, motivating a search for ASD in other contexts. Recently, ASD was found to hold for square fishnet integrals in a strongly-deformed version of planar $\mathcal{N}=4$. 
In this paper, we investigate the properties of polynomials in ladder integrals under the ASD transformation that leaves the square fishnet invariant. Our basic building blocks are determinants of certain matrices of ladder integrals, that are labeled by partitions or Young diagrams, and automatically obey the Steinmann relations.  We employ integrability-based representations of such determinants, which are essentially matrix-model integrals, and the theory of symmetric polynomials to find many more ASD quantities -- linear combinations of determinants whose coefficients are expressed using the combinatorics of partitions. As a byproduct, we provide an alternate proof that square fishnet integrals obey ASD.}

\begin{document}
\hypersetup{pageanchor=false}
\maketitle
\hypersetup{pageanchor=true}

\section{Introduction}
\label{sec:intro}

Antipodal self-duality (ASD) is a remarkable, but poorly understood, property of planar ($\mathcal{N}=4$) supersymmetric Yang-Mills theory (pMSYM). ASD relates an amplitude or form factor to itself, after performing a suitable kinematic map and reversing the letters of the symbol.  More precisely, the latter operation is the action of the antipode of the Hopf algebra of (the de Rham analogs of) multiple polylogarithms~\cite{Goncharov2001}.  However, to date ASD has been seen in just one process in pMSYM, namely the four-point form factor of the chiral stress tensor operator ${\rm tr}\phi^2$, where it holds, to at least three loops, when the four momenta in the process live in a three-dimensional (3D) subspace~\cite{Dixon:2022xqh}.
The 3D kinematics constitute a four-parameter kinematical space. 
Under the ASD kinematic map, two particular limits of the kinematics are mapped into each other.
(A cluster algebra interpretation of this map has been provided recently in ref.~\cite{He:2026kdq}.)
In one limit, when two particles become collinear, the (finite remainder of the) four-point form factor goes smoothly into the three-point form factor.  In the other limit, when three particles become collinear, the four-point form factor goes smoothly into the  six-gluon, maximally helicity violating (MHV) amplitude~\cite{Bern:2008ap,Dixon:2022xqh}. 
Because these two regions are mapped into each other, ASD for the four-point form factor {\it explains} the previously-observed~\cite{Dixon:2021tdw} antipodal duality (AD) between the three-point form factor and the six-point MHV amplitude.  This AD has been verified to eight loops~\cite{Dixon:2022rse,Dixon:2023kop}, and very recently at nine loops with the aid of AI~\cite{Anthropic9loops,Song9loops}. 
Many aspects of AD have been verified for the full antipode map, beyond the symbol.  In contrast, for the four-point ASD, only one beyond-the-symbol aspect has been explicitly checked at two loops~\cite{Dixon:2024yvq}.

Given that there is really only one example of ASD in pMSYM, and that
its physical origin is unknown, it is worth searching for more examples in neighboring theories.  Conformal fishnet theory~\cite{Gurdogan:2015csr,Caetano:2016ydc} is a strong deformation of pMSYM that sacrifices unitarity while preserving integrability~\cite{Chicherin:2017cns,Gromov:2017cja,Chicherin:2017frs} -- at least for fishnet graphs (including the ones considered in ref.~\cite{Basso:2017jwq}) where all external legs point ``outward''~\cite{Beisert:2025mtn}. 
One version of fishnet theory contains two complex scalars, and no fermions or gluons. Certain four-point, point-split single trace correlation functions, or {\it rectangular fishnets}, consist of single Feynman integrals that can be computed to all orders~\cite{Basso:2017jwq,Derkachov:2019tzo,Derkachov:2020zvv,Basso:2021omx} as determinants of {\it ladder integrals}~\cite{Usyukina:1992jd,Usyukina:1993ch}.
Recently it was pointed out by two of the authors that one class of these observables, the {\it square fishnets}, possesses ASD~\cite{Dixon:2025zwj}.

The purpose of this paper is to search for generalizations of square fishnets that also possess ASD.  A few initial examples were found in ref.~\cite{Dixon:2025zwj}, but here we will go much further.
The class of functions we will search through consists of polynomials in the ladder integrals, which also obey the Steinmann constraints~\cite{Steinmann,Steinmann2} on double discontinuities. We ask that linear combinations of these quantities obey ASD, for the same choice of kinematic map that works for the square fishnets. All the solutions we find are linear combinations of certain determinants appearing in large $R$-charge correlators~\cite{Coronado:2018cxj,Caron-Huot:2021usw}.
These determinants can be labelled by {\it partitions}, which are non-increasing, finite lists of integers.   The integers describe the removal of rows from the matrix whose determinant gives the square-fishnet integral.  Partitions are in one-to-one correspondence with Young diagrams, and they play an essential role in the theory of symmetric functions (polynomials).
We will see that the combinatorial quantities appearing in the ASD linear combinations all arise naturally in the theory of symmetric functions.  This observation can be understood from a dual representation of the fishnet integrals~\cite{Basso:2021omx}. 

In a little more detail, the $p$-loop ladder integrals are~\cite{Usyukina:1992jd,Usyukina:1993ch}
\begin{equation}
\label{eq:ladder_sum_def}
    f_p(z,\bar{z}) = \sum_{j=p}^{2p} \frac{(p-1)! j! \, {\cal L}^{2p-j}}{(j-p)! (2p-j)!} \, \left[ {\rm Li}_j(z) - {\rm Li}_j(\bar{z}) \right] \,,
\end{equation}
where $(z,\bar{z})$ are complex coordinates for the two independent cross ratios in a conformal four-point function, ${\cal L} \equiv -\log(z\bar{z})$, and ${\rm Li}_j(z) = \sum\limits_{k=1}^\infty \frac{z^k}{k^j}$ is the classical polylogarithm of weight $j$.  The logarithm has weight 1, so $f_p$ has weight $2p$.

The ASD transformation $\bfSh$ of the ladder integrals that was found~\cite{Dixon:2025zwj} to leave the square fishnet integrals invariant maps the ladder integrals $f_p \to \bfSh(f_p) \equiv \tilde{f}_p$, where $\tilde{f}_p$ is a binomial transform of $f_p$:
\begin{equation}
\label{eq:f_transf}
    \tilde{f}_p = \bfSh(f_p)= \sum_{k=0}^{p-1} \binom{p-1}{k} L^k f_{p-k} \,,
\end{equation}
with \begin{equation}
\label{eq:L_def}
    L \equiv \log z \times \log \bar{z} \,.
\end{equation}
Strictly speaking, the antipode map acts on de Rham analogs of polylogarithms, which can be understood as (motivic) polylogarithms modulo $i\pi$~\cite{Brown:2015ylf}. We will not make a distinction here, as we are mainly interested in the implications of the transformation~\eqref{eq:f_transf} on functions.

Ladder integrals are building blocks for other interesting quantities. The main result of ref.~\cite{Basso:2017jwq} is that rectangular $m \times n$ fishnet correlators are given, up to a simple prefactor, by the following determinant of an $m\times m$ matrix of ladder integrals:
\begin{equation}
\label{eq:BD_det}
\phi_{m,n} = \det_{1 \leq i,j \leq m} f_{n-m+i+j-1} \,,
\end{equation}
where $n\geq m$.   To be more precise, the full $m\times n$ fishnet correlator is the split-trace four-point function
\begin{equation} 
G_{m,n} \equiv \langle {\rm Tr} [ X^n(x_1) Z^m (x_3) \bar{X}^n(x_2) \bar{Z}^m(x_4) ] \rangle
\label{eq:Gmndef}
\end{equation}
in the fishnet theory with Lagrangian~\cite{Gurdogan:2015csr}
\begin{equation}
\label{eq:Lagr}
{\cal L} = {\rm Tr} [ X (-\Box) \bar{X} + Z (-\Box) \bar{Z} - (4\pi g)^2 XZ\bar{X}\bar{Z} ] \,.
\end{equation}
The four-point function depends essentially only on two conformally-invariant cross ratios, which can be parametrized in terms of the complex variable $z$ as
\begin{equation}
\label{eq:crossratiodefn}
    \dfrac{x_{14}^2 x_{23}^2}{x_{12}^2 x_{34}^2} = \dfrac{z\zb}{(1-z)(1-\zb)} \,,\quad \dfrac{x_{13}^2 x_{24}^2}{x_{12}^2x_{34}^2}= \dfrac{1}{(1-z)(1-\zb)} \,,
\end{equation}
where $x_{ij}^2=(x_i-x_j)^2$.
This function is related to $\phi_{m,n}$ by
\begin{equation}
    G_{m,n}= \dfrac{g^{2mn}}{x_{12}^{2n}x_{34}^{2m}} \left[\dfrac{(1-z)(1-\zb)}{z-\zb} \right]^m\dfrac{\phi_{m,n}(z,\bar{z})}{\mathcal{N}_{m,n}}\,,
\end{equation}
where the normalization constant $\mathcal{N}_{m,n}$ is given by
\begin{equation}
\label{eq:Nmn}
   \mathcal{N}_{m,n} = \prod_{k=0}^{2m-1}(n-m+k)!\,,
\end{equation}
and $g$ is the coupling in eq.~\eqref{eq:Lagr}. 

Using the ladder transformation formula~\eqref{eq:f_transf}, one can show~\cite{Dixon:2025zwj} that the square fishnets with $n=m$, $\phi_{m,m}(z,\zb)$, are self-dual under the antipode transformation, i.e.~$\bfSh(\phi_{m,m}) = \phi_{m,m}$. (We will present an alternate proof in this paper.)  The determinantal structure prompts an immediate generalization to a more general set of objects:
\begin{equation}\label{eq:Nlambda}
N_{\lambda}^{(m)} = \det_{1 \leq i,j \leq m} f_{m+\lambda_i+j-i} \ ,
\end{equation}
which are now labelled by partitions $\lambda$, as well as by the matrix size $m$ \cite{Coronado:2018cxj}. (We have suppressed the dependence on $(z,\bar{z})$.) A partition is a non-increasing, finite sequence of positive integers. The elements \eqref{eq:Nlambda} are defined to be  zero if the number of elements in $\lambda$ is greater than $m$. Note that the empty partition $\lambda\equiv0$ leads to the $m\times m$ matrix with elements $f_{m+j-i}$.  This matrix may be re-indexed by swapping top and bottom rows, i.e.,~by letting $i \to m+1-i$, in order to recover the square fishnet matrix $f_{i+j-1}$ in eq.~\eqref{eq:BD_det} with $n=m$. Hence the determinants are identical, up to an overall sign factor of $(-1)^{m(m-1)/2}$ from the row swaps.  Similarly, the rectangular fishnet determinant is recovered by taking $\lambda = [n-m,n-m,\ldots,n-m]$ ($m$ copies of $(n-m)$).   The matrices in eq.~\eqref{eq:BD_det} are called Hankel matrices, while those in eq.~\eqref{eq:Nlambda} are Toeplitz matrices, but their determinants are equal up to the sign factor.

Objects such as $N_{\lambda}^{(m)}$ have already appeared as the basis for the expansion of an {\it octagon} large $R$-charge correlator~\cite{Coronado:2018cxj,Caron-Huot:2021usw}.  Like the fishnet four-point function, the octagon is a planar object with a dual scattering interpretation, for which the Steinmann relations~\cite{Steinmann,Steinmann2} imply the vanishing of a certain double discontinuity.  Combined with conformal symmetry, the condition turns out to be~\cite{Basso:2017jwq} that 
\begin{equation}
\label{eq:Stzeq1}
 {\rm Disc}_{z=1} \Big[ {\rm Disc}_{z=1} N_{\lambda}^{(m)} \Big] = 0\,.
 \end{equation}
The single discontinuity around $z=1$ of the ladder integral is~\cite{Basso:2017jwq}
\begin{equation}
\label{eq:Disc1fp}
    {\rm Disc}_{z=1} f_p = - 2\pi i \, \log \left(\frac{z}{\bar{z}}\right) \, (-L)^{p-1} \,.
\end{equation}
Thus any row in a matrix of the form $f_{m+\lambda_i+j-i}$, as in eq.~\eqref{eq:Nlambda}, with the simple dependence of the index on $j$, has a discontinuity whose $j$-dependence is proportional to $(-L)^j$. The vanishing of determinants with proportional rows then implies eq.~\eqref{eq:Stzeq1}, as shown in ref.~\cite{Coronado:2018cxj}. It is important for the Steinmann condition that only the partition $\lambda_i$ appears, and not also $\lambda_j$. (Dependence on $\lambda_j$ has been considered in other contexts~\cite{bumpdiaconis2002toeplitz}.)

The $N_\lambda^{(m)}$ determinants defined in eq.~\eqref{eq:Nlambda} are also ubiquitous in the study of random matrix theory and integrability \cite{miwa2000solitons,DiFrancesco:1993cyw,Morozov:1994hh,deift2000orthogonal}. The main output of this paper is that these quantities are natural building blocks of ASD functions. First, we show that the $N_\lambda^{(m)}$ transform under the antipode according to generalized binomial expansions. Next, using the antipode transformation formula, we find that one can construct ASD functions as linear combinations of eq.~\eqref{eq:Nlambda}. The general formula turns out to be relatively compact; hence, we present it here.   For clarity, we will generally suppress the dependence on $(z,\bar{z})$ in the rest of this article.

While the $N_{\lambda}^{(m)}$ are defined for all partitions, the ASD functions that we find, $Q_{\lambda}^{(m)}(z,\zb) \equiv Q_\lambda^{(m)}$, are  labelled by \emph{nuclear} partitions -- that is, partitions $\lambda$ with a restriction that all $\lambda_i > 1$ (see section~\ref{sec:ShiftInvariant}). For each such partition, the ASD function is expressed as a sum over subpartitions $\nu$ of $\lambda$ (denoted by $\nu \subseteq \lambda$) having the same length as $\lambda$:
\definecolor{crimson}{rgb}{0.86, 0.08, 0.24}
\definecolor{bleudefrance}{rgb}{0.19, 0.55, 0.91}
\definecolor{green(html/cssgreen)}{rgb}{0.0, 0.5, 0.0}
\definecolor{azure(colorwheel)}{rgb}{0.0, 0.5, 1.0}
\begin{equation}\label{eq:QBasis2}
Q_{\lambda}^{(m)} = \sum_{\substack{\nu \subseteq\lambda \\ l(\nu)=l(\lambda)}} \dfrac{(-1)^{|\lambda|-|\nu|} {\color{crimson}f^{\lambda/\nu}}}{\color{azure(colorwheel)}\sfC_{{\nu}\cup (1^{|\lambda|-|\nu|})}(m) } {\color{green(html/cssgreen)}N_{{\nu}\cup (1^{|\lambda|-|\nu|})}^{(m)}} .
\end{equation}
Here we employ the following notation: For a partition $\lambda=[\lambda_1,\lambda_2,\ldots,\lambda_{l(\lambda)}]$, by definition the positive-integer components $\lambda_i$ are non-increasing, $\lambda_1 \geq \lambda_2 \geq \ldots \geq \lambda_{l(\lambda)}>0$. The number of non-zero components of the partition is its length $l(\lambda)$. The sum of all components is given by $|\lambda|=\lambda_1+\cdots+\lambda_{l(\lambda)}$. The partition ${\nu}\cup (1^{|\lambda|-|\nu|})$ is obtained by appending to the partition $\nu$ a string of $1$'s of length ${|\lambda|-|\nu|}$. Note that the sum of the components of the resulting partition ($|\nu| + |\lambda|-|\nu|$) is equal to that of $\lambda$.  

There are four components to formula~\eqref{eq:QBasis2}. The first component is the ladder determinant $\color{green(html/cssgreen)}N_{{\nu}\cup (1^{|\lambda|-|\nu|})}^{(m)}$; it is the only component on the right-hand side that depends on $(z,\bar{z})$. The (constant) coefficients in the sum are a product of three factors.  The first factor is just a sign depending on the number of appended 1's, ($-1)^{|\lambda|-|\nu|}$. The second factor is denoted by $\color{azure(colorwheel)}\sfC_{{\nu}\cup (1^{|\lambda|-|\nu|})}(m) $; it is a product of factors that are linear in $m$, which are given explicitly in eq.~\eqref{eq:ContentProduct}. The third constant factor is ${\color{crimson}f^{\lambda/\nu}}$; it is just an integer which counts the number of ways to obtain the partition $\lambda$ from the partition $\nu$ by increasing the components of $\nu$ successively by one, in such a way that any intermediate step is itself a partition. The number ${\color{crimson}f^{\lambda/\nu}}$ is also the number of paths between $\lambda$ and $\nu$ in the Young's lattice of partitions. See appendix~\ref{sec:appendix}, where a closed formula~\eqref{eq:flmdetapp} is provided. (It is also given in eq.~\eqref{eq:flmdet}.)

Consider for example, the ASD combination associated with the nuclear partition $[2,2]$.  Equation~\eqref{eq:QBasis2} evaluates to:
\begin{equation}
    \begin{split}
         Q_{[2,2]}^{(m)}\ =& \ \dfrac{ {\color{green(html/cssgreen)}N_{[2,2]}^{(m)}}}{\color{azure(colorwheel)} m^2(m+1)(m-1)}-\dfrac{ {\color{green(html/cssgreen)}N_{[2,1,1]}^{(m)}}}{\color{azure(colorwheel)}m(m+1)(m-1)(m-2)} + \\
         &+ \dfrac{ {\color{green(html/cssgreen)}N_{[1,1,1,1]}^{(m)}}}{\color{azure(colorwheel)}m(m-1)(m-2)(m-3)} \,.
    \end{split}
 \end{equation}
The subpartitions of $\lambda = [2,2]$ with the same length are $[2,2]$, $[2,1]$, and $[1,1]$.  After appending strings of $1$'s until the sum of components equals $|\lambda| = 4$, the resulting partitions are $[2,2]$, $[2,1,1]$, $[1,1,1,1]$; this list determines which $\color{green(html/cssgreen)}N_{{\nu}\cup (1^{|\lambda|-|\nu|})}^{(m)}$ terms appear on the right-hand side. The ${\color{crimson}f^{\lambda/\nu}}$ quantities in this case are all equal to one, and only one sign is negative.
For more examples and a detailed discussion see section \ref{sec:with_examples}.

The rest of the paper is organized as follows. In section~\ref{sec:dualrep} we introduce an integral representation of rectangular fishnet correlators and map the antipodal transformation to a shift operator on symmetric functions. Section~\ref{sec:ShiftInvariant} constructs a basis in the space of shift-invariant symmetric functions and the corresponding ASD ladder polynomials. In section~\ref{sec:with_examples} we provide more examples with detailed explanations.  Section~\ref{sec:proof_invariance} proves that a certain polynomial $W_\lambda$ in the symmetric variables is invariant under the shift, while section~\ref{sec:m_too_small} describes the special, degenerate case where $m<|\lambda|$.  Finally, in section~\ref{sec:Conclusions} we recap and provide some avenues for further research. Excerpts from the theory of symmetric functions and combinatorics of partitions are collected in appendix~\ref{sec:appendix}.

\section{Antipodal transformations and dual representations of fishnet integrals}\label{sec:dualrep}

In this section we review the formulation of fishnet integrals in terms of a dual matrix model. It is known that this dual formulation provides a framework for Basso-Dixon-type determinant representations~\cite{Basso:2017jwq}. We will show that this representation allows one to incorporate antipodal transformations in a very natural way.

\subsection{The dual representation of fishnet integrals}

We start by reviewing the dual representation of ladder integrals, followed by the rectangular fishnet integrals~\cite{Basso:2021omx}. 

Ladder integrals have an integral representation which is the key ingredient in our construction below.  To obtain it, following ref.~\cite{Basso:2021omx}, one first represents the polylogarithm as follows:
\begin{equation}
\label{eq:Linintegralrep}
    \Li_n(z) =\dfrac{z}{(n-1)!} \int\limits_{0}^\infty \dfrac{\mathrm{d}r\,r^{n-1}}{e^{r}-z} \,.
\end{equation}
Then, the sum in eq.~\eqref{eq:ladder_sum_def} can be taken under the integral sign.  Letting $r=\tilde{x}+\sigma$ and using a combinatorial identity to perform the sum, one obtains for the $p$-loop ladder integral:
\begin{equation}
\label{eq:ladderdual}
  f_p(z,\zb) =   \dfrac{z-\zb}{2\sqrt{z\zb}} \int_{|\sigma|}^{\infty} \dfrac{\mathrm{d}\tx\, \tx\,(\tx^2-\sigma^2)^{p-1}}{\cosh \tfrac{1}{2}{\left(\tx+\varphi\right)}\cosh\tfrac{1}{2}{\left(\tx-\varphi\right)}} \,,
\end{equation}
where we introduced the variables $\sigma$ and $\phi$ via
\begin{equation}
\label{eq:sigmaphidef}    
z= -\exp(\sigma+\varphi)\,, \qquad \zb= -\exp(\sigma-\varphi)\,.
\end{equation}

The dual representation for the ladder integrals can be lifted to the whole fishnet four-point function using the Cauchy-Binet-Andr\'{e}ief formula, as explained in ref.~\cite{Basso:2021omx}. One then obtains for the rectangular fishnet:
\begin{equation}
\label{eq:phimndual}
    \phi_{m,n}(z,\zb) = \dfrac{1}{m!} \left( \dfrac{z-\zb}{2\sqrt{z\zb}} \right)^m \int_{|\sigma|}^{\infty}  \prod_{i=1}^m \frac{\mathrm{d} \tx_i\, \tx_i\, (\tx_i^2-\sigma^2)^{n-m}}{\cosh \frac{1}{2}\left(\tx_i+\varphi\right) \cosh \frac{1}{2}\left(\tx_i-\varphi\right)} \prod_{i<j}^m\left(\tx_i^2-\tx_j^2\right)^2\,.
\end{equation}
Note that the ladders are recovered as $f_p = \phi_{1,p}$.

Integrals similar to the one in eq.~\eqref{eq:phimndual} appear in the theory of random matrix models. Unitarily invariant random matrix models can be brought into eigenvalue form, in which case they define a distribution on the set of random eigenvalues $\bfx= (x_1,\ldots, x_m)$.  Moments of this distribution are integrals of symmetric polynomials $g(\bfx)$ of the form:
\begin{equation}\label{eq:matrix_integral}
    \ev{g(\bfx)} = \int_{\Gamma} \prod_{i=1}^m \mathrm{d}x_i\, w(x_i)\, \prod_{i<j}(x_i-x_j)^2 g(\bfx)\,.
\end{equation}
The function $w(x)$ depends on a single eigenvalue weight and specifies a particular random matrix model.  The model also depends on the range of integration, or contour, $\Gamma$, which is chosen so that moments of the distribution $w(x)$ exist. The precise definition of $\Gamma$ is not important for the general arguments below. The bilocal product is the square of the Vandermonde determinant, which arises from integrating out the `angular' variables in the full matrix integral. It is usually denoted by
\begin{equation}\label{eq:Vander}
    \Delta(\bfx)= \prod_{1\leq i<j\leq m}(x_i-x_j) =\det_{i,j}\left(  x_j^{m-i}\right) \,.
\end{equation}

It is now possible to interpret the dual integral in eq.~\eqref{eq:phimndual} as a matrix model integral of the form given in eq.~\eqref{eq:matrix_integral}. 
To this effect, we change integration variables in eq.~\eqref{eq:phimndual} to $x_i = \tx_i^2-\sigma^2$, and we define the fishnet matrix model as
\begin{equation}
\label{eq:g_exp}
\ev{g({\bf x})}_{\mathrm{F}}^{(m)}= \frac{(-1)^{m(m-1)/2}}{m!}\int_{0}^{\infty}  \prod_{i=1}^m \mathrm{d}x_i\, w_{\rm F}(x_i)\, \prod_{i<j}(x_i-x_j)^2 \, g(\bfx)\,,
\end{equation}
with
\begin{equation}
   w_{\rm F}(x) = \left( \dfrac{z-\zb}{2\sqrt{z\zb}} \right)\cdot \dfrac{1}{\cosh\sqrt{x+\sigma^2}+\cosh\varphi} \,.
\end{equation}
It is well known that expectation values of certain types of functions in one-matrix models with rather generic potentials can be written in determinant form~\cite{forrester2010log,forrester2019meet}.

The specific form of the square fishnet determinant suggests that in our case the relevant functions are \emph{Schur polynomials}. They are symmetric polynomials of $\bfx=(x_1,\ldots,x_m)$ that  are labeled by partitions of some integer $n$, i.e., by a sequence $\lambda = [\lambda_1,\ldots,\lambda_r]$ such that $\lambda_i\ge \lambda_{i+1}>0$ and $r\le m$. $l(\lambda)=r$ is the \emph{length} of the partition.
The Schur polynomials can be defined via the determinant formula:
\begin{equation} \label{eq:Schur}
    s_\lambda(\bfx) = \dfrac{\det x_j^{m+\lambda_i-i}}{\Delta(x)} \,.
\end{equation}
Schur polynomials  are homogeneous of degree $|\lambda|=\sum_{i=1}^r\lambda_i$, and they provide a basis for the ring of symmetric polynomials. We recall more properties of Schur polynomials together with some other facts about symmetric functions in appendix \ref{sec:appendix}. Using the Cauchy-Binet-Andréief identity, one can rewrite the expectation values of Schur polynomials in a determinant form:
\begin{equation}\label{eq:SchurDet}
\begin{split}
        \ev{s_\lambda(\bfx)}_{\!\rm F}^{\!(m)} = &\ \dfrac{1}{m!}\int_{0}^{\infty} \prod_{i=1}^m
\mathrm{d}x_i  \,
\det\left( w_\mathrm{F}(x_i)\,x_i^{j-1} \right)\det x_j^{m+\lambda_i-i} 
\\=&\ \det{\int_{0}^{\infty}\mathrm{d}x\, w_\mathrm{F}(x)\, x^{m+\lambda_i+j-i-1}}\,,
\end{split} 
\end{equation}
where all determinants are over $1\leq i,j \leq m$, and we used the second expression for the Vandermonde determinant in eq.~\eqref{eq:Vander}.

We can observe that the determinants appearing in the four-point fishnet correlation functions of eq.~\eqref{eq:BD_det} almost match the form in eq.~\eqref{eq:SchurDet}. The two expressions are different in that the matrix of ladder integrals in eq.~\eqref{eq:BD_det} is of Hankel form and that in eq.~\eqref{eq:SchurDet} is of Toeplitz form. The determinant can be written in either form. We therefore reinterpret the result of ref.~\cite{Basso:2021omx} as follows: rectangular fishnet correlation functions are given by expectation values of rectangular Schur polynomials:
\begin{equation}
    \phi_{m,n}(z,\zb)= (-1)^{m(m-1)/2} \, \ev{s_{[(n-m)^m]}(\bfx)}_{\!\rm F}^{\!(m)}\,.
\end{equation}
Here $[(n-m)^m]$ is the notation for a partition of the form $[\underbrace{(n-m),\ldots,(n-m)}_m]$, and the sign factor accounts for the switch between Hankel and Toeplitz conventions. In particular, for the ($m=1$) ladder integrals $f_n(z,\bar{z}) = \phi_{1,n}(z,\bar{z})$ we find 
\begin{equation}
    f_n(z,\zb)= \ev{s_{[n-1]}(x)}_{\!\rm F}^{\!(1)} = \int_{0}^{\infty}\mathrm{d}x\, w_\mathrm{F}(x)\, x^{n-1}\,.
\end{equation}
For other partitions we get generalizations of the rectangular fishnet integrals. These are precisely the ladder determinants $N_{\lambda}^{(m)}(z,\zb)$ introduced in ref.~\cite{Coronado:2018cxj}, and we have: 
\begin{equation}\label{eq:NSchur}
    N_{\lambda}^{(m)}(z,\zb) = \ev{s_{\lambda}(\bfx)}_{\!\rm F}^{\!(m)} = \det f_{m+\lambda_i+j-i}(z,\zb)\,.
\end{equation}
From here on we suppress the dependence on $(z,\bar{z})$ for clarity and simply write $  N_{\lambda}^{(m)}$ for the determinants and similar functions. 
In ref.~\cite{Coronado:2018cxj} it was conjectured that the determinants $N_{\lambda}^{(m)}$ form a basis for all polynomials in ladder integrals $f_p$ that satisfy the Steinmann conditions~\cite{Steinmann,Steinmann2}. Since the Schur polynomials form a basis for all symmetric polynomials, eq.~\eqref{eq:NSchur} allows us to identify the linear subspace generated by the $N_{\lambda}^{(m)}$ with the vector space of symmetric polynomials in $m$ variables. In particular, any question related to polynomials in ladder integrals that satisfy Steinmann conditions must have an analogue in the space of symmetric polynomials.

\subsection{Antipodal transformations in the dual representations}

We now argue that the antipodal transformation $\bfSh$ in eq.~\eqref{eq:f_transf} has a very simple description in eq.~\eqref{eq:g_exp}, the dual matrix model. Consider a symmetric polynomial $g(\bfx)$ in $m$ variables and a formal variable $L$. We define an operator on the vector space of symmetric polynomials by
\begin{equation}
\mathbb{S}_L(g)(\bfx) = g(\bfx+L)\,,
\end{equation}
where we introduced the shorthand notation
\begin{equation}
    \bfx+L = (x_1+L,\ldots,x_m+L)\,.
\end{equation}
Note that, if $g$ is a symmetric polynomial in $\bfx$ with constant coefficients, then $\mathbb{S}_L(g)(\bfx)$ is also a symmetric polynomial in $\bfx$, but with coefficients that are themselves polynomials in $L$. Alternatively, we can think of $\mathbb{S}_L(g)(\bfx)$ as a polynomial in $L$ with coefficients that are symmetric polynomials in $\bfx$.
Our goal is to show that for $L=\log z\,\log\zb$, the operator $\mathbb{S}_L$ is the analogue of the antipodal map $\bfSh$ acting on the functions $N_{\lambda}^{(m)}$, but pulled back to the space of symmetric polynomials. More precisely, we will show that
\begin{equation}\label{eq:pol_S_def}
\ev{\mathbb{S}_L(g)(\bfx)}^{\! (m)}_{\!\rm F} = \bfSh\ev{g(\bfx)}^{\! (m)}_{\!\rm F}\,,\qquad L=\log z\,\log\zb\,.
\end{equation}

The proof of this statement proceeds by direct computation. Any symmetric polynomial $g(\bfx)$ can be expanded in the basis of Schur polynomials, hence it is enough to prove eq.~\eqref{eq:pol_S_def} when $g$ is a Schur polynomial. We start by showing it for Schur polynomials associated with  the ladder integrals $ f_n(z,\zb) = \ev{s_{[n-1]}(x)}_{\!\rm F}^{\!(1)}$. Using the binomial expansion of $(x+L)^{n-1}$, we have
\begin{equation}\label{eq:shift_ladder}
\begin{split}
    \ev{\mathbb{S}_{L}(s_{[n-1]})(x)}^{\!(1)}_{\!\rm F} &\,= \ev{s_{[n-1]}(x+L)}^{\!(1)}_{\!\rm F}\\
    &\,= \int_{0}^{\infty} \mathrm{d} x\, w_{\rm F}(x) \left(x+L \right)^{n-1}\\
    &\,=\sum_{k=0}^{n-1}\binom{n-1}{k}\,L^{k}\,\ev{s_{[n-k-1]}(x)}^{\!(1)}_{\!\rm F}\\
     &\,=\sum_{k=0}^{n-1}\binom{n-1}{k}\,L^{k}\,f_{n-k}\\
     &\, =\bfSh\left(f_n \right)\,.
\end{split}\end{equation}
Comparing the last expression to eq.~\eqref{eq:f_transf}, we immediately see that eq.~\eqref{eq:pol_S_def} holds for $g=s_{[n-1]}$. 

Next, let us consider $m> 1$.
We may then use eq.~\eqref{eq:NSchur} and apply the shift to individual moments. More precisely, we have the following chain of equalities:
\begin{equation}
 \begin{split}
        \ev{ \mathbb{S}_{L}(s_{\lambda})(\bfx) }^{\!(m)}_{\!\rm F}&=\ev{ s_{\lambda}(\bfx+L) }^{\!(m)}_{\!\rm F}\\&= \det \int_{0}^{\infty} \mathrm{d}x\, w_{\rm F}(x) \left(x+L \right)^{m+\lambda_i+j-i-1}
        \\
        &=\det\bfSh\left(f_{m+\lambda_i+j-i} \right) \\
        &= \bfSh\left( N_{\lambda}^{(m)} \right)\,.
 \end{split}
\end{equation}
Hence, eq.~\eqref{eq:pol_S_def} holds for all partitions $\lambda$, which finishes the proof.  

In conclusion, the action of the antipodal transformation $\bfSh$ on any generalized fishnet determinant $N_\lambda^{(m)}$ can be deduced from the action of the shift transformation $\bfx \rightarrow \bfx + L$ on the associated Schur polynomial $s_\lambda(\bfx)$.

The action of the operator $\mathbb{S}_L$ implementing this transformation can be described using the following property of Schur polynomials. The Schur polynomial with a shifted argument is still a symmetric function. Hence it can be reexpanded back into Schur polynomials, by keeping track of the degree of homogeneity:
  \begin{equation}\label{eq:shift_s}
        s_\lambda(\bfx+a) = \sum_{\mu\,:\, |\mu|\leq|\lambda|} a^{|\lambda|-|\mu|}\,b_{\lambda\mu} \,s_\mu(\bfx) \,,
    \end{equation}
where the sum is over partitions, or equivalently Young diagrams, whose total number of boxes is less than or equal to that of $\lambda$.  The key property is that the coefficients $b_{\lambda\mu}$ can be found in closed form. First, the coefficients are non-vanishing only for partitions $\mu$ that are included in $\lambda$ (see appendix \ref{sec:appendix}),
\begin{equation}
    \mu \subseteq \lambda \,  \quad\Leftrightarrow\quad \mu_i \leq \lambda_i \ \hbox{for all $i$ such that}\ 1\leq i \leq l(\mu) \leq l(\lambda)\,.
\end{equation}

For such partitions we have \cite{Macdonald:book}:
 \begin{equation}
 \label{eq:b_formula}
        b_{\lambda\mu}= \left\{ \begin{split}
            &\dfrac{f^{\lambda/\mu}}{(|\lambda|-|\mu|)!} \dfrac{\sfC_\lambda(m)}{\sfC_{\mu}(m)} \,,& \quad\mu \subseteq \lambda\,,
            \\
            &0 \, ,& \quad \mu \not\subseteq \lambda \,.
        \end{split} \right.
    \end{equation}
Here we used two standard quantities from the theory of symmetric functions. First, $f^{\lambda/\mu}$ is the number of standard Young tableaux for the skew-shape $\lambda/\mu$, see appendix~\ref{sec:appendix}. It is given by the generalized hook length formula, explicitly (cf.,~e.g.,~corollary 7.16.3 in ref.~\cite{stanley2023enumerative2}):
    \begin{equation}
    \label{eq:flmdet}
        f^{\lambda/\mu} = (|\lambda|-|\mu|)!  \,\det \left( \dfrac{1}{(\lambda_i-\mu_j+j-i)!} \right) \,,
    \end{equation}
    where $1/k!$ is defined to be $0$ if $k<0$, for $k = \lambda_i - \mu_j + j - i$. The quantity $\sfC_\lambda(m)$ is the so-called \emph{content product}:
\begin{equation}\label{eq:ContentProduct}
        \sfC_\lambda(m) = \prod_{(i,j)\in \lambda} (m+j-i)  = \prod_{i=1}^{l(\lambda)}\prod_{j=1}^{\lambda_i} (m+j-i)\,,
    \end{equation}  
where $(i,j)$ are coordinates of boxes in the Young diagram. We explain both quantities in greater detail in appendix~\ref{sec:appendix}. The content product $\sfC_\lambda(m)$ can be computed diagrammatically by putting an $m$ in the upper left box $(1,1)$ of the Young diagram for $\lambda$, an $m+1$ ($m-1$) in the box to the right of (below) that box, and so on. The content ratio $\sfC_\lambda(m)/\sfC_\mu(m)$ in eq.~\eqref{eq:b_formula} can be computed from the boxes in the Young diagram for $\lambda$ that are {\it not} in the Young diagram for $\mu$.  These are also the boxes of the skew diagram $\lambda/\mu$, see appendix~\ref{sec:appendix}.

Putting everything together, we see that the action of the operator $\mathbb{S}_L$ on Schur polynomials can be entirely described in terms of quantities from the theory of symmetric polynomials,
\begin{equation}\label{eq:pol_S_action}
        \mathbb{S}_L\left( s_{\lambda}\right)(\bfx)  = \sum_{\mu \subseteq \lambda} \dfrac{f^{\lambda/\mu} L^{|\lambda|-|\mu|}}{(|\lambda|-|\mu|)!} \dfrac{\sfC_\lambda(m)}{\sfC_{\mu}(m)} \, s_{\mu}(\bfx)\,.
    \end{equation}

    We can use eq.~\eqref{eq:pol_S_action} to obtain a closed form for the action of the antipodal map $\bfSh$ on the generalized fishnet integrals $N_{\lambda}^{(m)}$. Indeed, 
    taking matrix model expectation values on both sides of eq.~\eqref{eq:pol_S_action} and using eq.~\eqref{eq:pol_S_def}, we obtain:
    \begin{equation}\label{eq:antipodal_transformation}
        \bfSh\left( N_{\lambda}^{(m)} \right) = \sum_{\mu \subseteq \lambda} \dfrac{f^{\lambda/\mu} L^{|\lambda|-|\mu|}}{(|\lambda|-|\mu|)!} \dfrac{\sfC_\lambda(m)}{\sfC_{\mu}(m)} \, N_{\mu}^{(m)}\,. 
    \end{equation}
    To illustrate this construction, we consider a couple of examples.

\paragraph{Example 1.} As a first example, let us discuss how we can recover the antipodal transformation of ladder integrals in eq.~\eqref{eq:f_transf} from the general formula in eq.~\eqref{eq:antipodal_transformation}.
For partitions of length one, $\lambda=[p-1]$ and $m=1$, we recover the ladder integrals:
\begin{equation}
    N_{[p-1]}^{(1)}=f_{p}\,.
\end{equation}
Then, the partitions $\mu$ appearing in the sum in eq.~\eqref{eq:antipodal_transformation}  also have length at most one, and $f^{[n]/[k]}=1$ if $k \leq n$. Hence, we find:
\begin{equation}
\label{eq:ladshift}
    \bfSh\left( N_{[p-1]}^{(1)} \right) = \sum_{k=0}^{p-1} L^{p-1-k}\, \binom{p-1}{k}\, N_{[k]}^{(1)}\,,
\end{equation}
where we used the fact that
\begin{equation}
    \dfrac{1}{(p-1-k)!} \dfrac{\sfC_{[p-1]}(1)}{\sfC_{[k]}(1)}= \binom{p-1}{k}\,.
\end{equation}
Relabelling $k \rightarrow p-1-k$ in eq.~\eqref{eq:ladshift}, we recover the transformation of ladder integrals from eq.~\eqref{eq:f_transf}.

\paragraph{Example 2.}
We now consider a partition $\lambda=[p-1]$ of length one and $m=2$. We have:
\begin{equation}
    N_{[p-1]}^{(2)}=f_2 f_{p+1}-f_1f_{p+2} \,.
\end{equation}
The partitions $\mu$ appearing in the sum in eq.~\eqref{eq:antipodal_transformation} again have length at most one. However, the coefficients are different compared to the case of ladder integrals:
\begin{equation}
    \bfSh\left( N_{[p-1]}^{(2)} \right) = \sum_{k=0}^{p-1} L^{p-1-k}\, \binom{p}{k+1}\, N_{[k]}^{(2)}\,.
\end{equation}

\paragraph{Example 3.}
 More structures appear for longer partitions. Consider the partition of length two $\lambda=[2,2]$, which corresponds to the Young diagram 
 \begin{equation}
\ytableausetup{boxsize=0.75em,aligntableaux = center}  \quad \ydiagram{2,2}
\end{equation}
We may express the generalized fishnet integral $N_{[2,2]}^{(m)}$ in terms of ladder integrals. For $m=2$ it is given by:
 \begin{equation}
     N_{[2,2]}^{(2)}=f_4^2-f_3f_5\,.
 \end{equation}
The action of the antipodal map for general $m$ can then be obtained from eq.~\eqref{eq:antipodal_transformation}:
\begin{equation}
 \begin{split}
        \bfSh\left(N_{[2,2]}^{(m)} \right) &= N_{[2,2]}^{(m)}+ m\, L\, N_{[2,1]}^{(m)}
    \\
    &+\dfrac{m(m-1)}{2}\,L^2\,N_{[2]}^{(m)}+\dfrac{m(m+1)}{2}\,L^2\,N_{[1,1]}^{(m)}\\
    &+\dfrac{2m(m+1)(m-1)}{6}\,L^3\,N_{[1]}^{(m)}+\dfrac{2m^2(m-1)(m+1)}{24}\, L^4 \,N_{\varnothing}^{(m)},
 \end{split}
\end{equation}
where $\varnothing = [0]$ is the empty partition.
Note that in this case we have some multiple paths: $f^{[2,2]/[1]} = f^{[2,2]/[0]} = 2$.

\section{ASD combinations of generalized fishnet integrals
}

In this section we present the main result of our paper, namely we present a closed expression for all combinations of generalized fishnet integrals $N_{\lambda}^{(m)}$ that are invariant under the action of the antipodal map $\bfSh$. More precisely, we are looking for linear combinations $Q$ of generalized fishnet integrals that satisfy
\begin{equation}
    \bfSh\big(Q\big) = Q\,.
    \end{equation}
In ref.~\cite{Dixon:2025zwj}, some of the authors have already shown that the square fishnet integrals $\phi_{n,n}$ are ASD, and have given a few more sporadic examples in terms of ladder integrals. We now extend this result, and we present a complete classification of all ASD combinations of generalized fishnet integrals $N_{\lambda}^{(m)}$.

From the results in the previous section, we know that every linear combination $Q^{(m)}$ of generalized fishnet integrals (self-dual or not, and for a given fixed choice of $m$) can be represented by a symmetric  polynomial $q$,
\begin{equation}
    Q^{(m)} = \ev{q(\bfx)}^{\!(m)}_{\!\rm F}\,.
\end{equation}
Equation~\eqref{eq:pol_S_def} then implies that the antipodal self-duality of $Q$ translates into a shift-invariance of the symmetric polynomial $q$,
\begin{equation}   \label{eq:shift-invariance}\mathbb{S}_L(q)(\bfx) = q(\bfx+L)=q(\bfx)\,.
\end{equation}
In other words, finding all ASD combinations $Q$ of generalized fishnet integrals reduces to finding all shift-invariant symmetric polynomials $q$, i.e., we want to find the subspace of all shift-invariant functions inside the space of all symmetric polynomials. Note that we have a basis for the full space of symmetric polynomials, given by the Schur polynomials, and we can always expand our symmetric polynomial in this basis,
\begin{equation}
    q(\bfx) =\sum_{\mu:|\mu| \leq \deg q(\bfx)} \alpha_{\mu}^{(m)}\, s_{\mu}(\bfx)\,,
\end{equation}
where the coefficients may depend on $m$. Our goal is then to find all coefficients $\alpha_{\mu}^{(m)}$ such that eq.~\eqref{eq:shift-invariance} holds. 

Before attacking that goal, we recall that the square fishnet integrals, $\phi_{m,m}(z,\bar{z})$, map to the empty partition, and to the trivial Schur polynomial $s_{\varnothing}(\bfx) = 1$, which obviously obeys eq.~\eqref{eq:shift-invariance}. Thus, in this framework, by exploiting the ladder shift transformation~\eqref{eq:shift_ladder}, and the dual representation~\eqref{eq:SchurDet}, it is trivial to prove that square fishnet integrals are antipodally self-dual.

\subsection{Shift-invariant symmetric polynomials}\label{sec:ShiftInvariant}

Before we present and prove our result, we first introduce a special class of partitions.

\paragraph{Nuclear partitions.}
Let $\cP(n)$ denote the set of all non-empty partitions of the positive integer $n$, and $p(n) = \#\cP(n)$ is its cardinality.
As already alluded to in the introduction, ASD combinations of fishnet integrals can be classified by partitions where no element is equal to one, or equivalently, whose associated Young diagram has no rows containing only a single box. 
 Such partitions have been studied in the literature~\cite{guy1958two,schneider2019nuclear,akandecomputational}, and they have appeared in the works of MacMahon and Sylvester \cite{macmahon1883seminvariants,sylvester1882subvariants}. They are called \emph{nuclear partitions} \cite{schneider2019nuclear}.\footnote{Sometimes these partitions are also referred to as \emph{non-unitary partitions}. As this term may lead to confusion in a physical context, we stick to nuclear.} We denote the subset of nuclear partitions of $\cP(n)$ by $\mathcal P^{\hat{1}}(n)$:
\begin{equation}
    \mathcal P^{\hat{1}}(n)  = \big\{ \lambda = [\lambda_1,\ldots,\lambda_r] \in \mathcal{P}(n) \, \big| \, \lambda_{i} > 1  \big\}\subset \mathcal{P}(n)\,.
\end{equation}
The total number of nuclear partitions of $n$ is given by~\cite{guy1958two}:
\begin{equation}
    p^{\hat{1}}(n)=p(n)-p(n-1)\,.
\end{equation}
Examples of the few nuclear partitions for $n\le 6$ are:
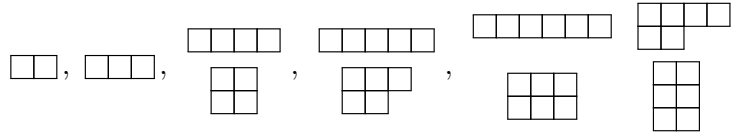
\begin{equation*}
     \ydiagram{2} \,, \ \ydiagram{3} \,,\  \begin{array}{c}
            \ydiagram{4}\vspace{0.1cm}\\
        \ydiagram{2,2}
    \end{array}
    \, , \ 
    \begin{array}{c}
            \ydiagram{5}\vspace{0.1cm}\\ 
        \ydiagram{3,2}
    \end{array}
     \, , \ 
    \begin{array}{cc}
            \ydiagram{6} & \hspace{0.2cm}\ydiagram{4,2}\vspace{0.15cm}\\ 
        \ydiagram{3,3} & \ydiagram{2,2,2}
    \end{array}
\end{equation*}
We will also need partitions associated with Young diagrams with row lengths greater than one, but less than or equal to $m$. We denote the set of such partitions by:
\begin{equation}
    \mathcal{P}^{\hat{1},m}(n) =  \big\{ \lambda \in \mathcal{P}(n) \, \big| \, \lambda_{i} > 1 \,,  
    \lambda_i \leq m\big\} \subseteq \cP^{\hat{1}}(n)\,.
\end{equation}

\paragraph{Nuclear partitions and shift-invariance.}
We know that a basis of all symmetric polynomials is given by the Schur polynomials, which are labeled by partitions. We now argue that shift-invariant symmetric polynomials are in one-to-one correspondence with nuclear partitions.
To see this, we go to ``center-of-mass'' variables, i.e., we choose new variables:
\begin{equation}\label{eq:COMchange}
    y_i= x_i - \bar{x} \, ,\qquad \bar{x} = \dfrac{1}{m}\sum_{i=1}^m x_i\,.
\end{equation}
It is easy to see that every symmetric function of the $\bf y$ variables is manifestly shift-invariant in the $\bf x$ variables. 

As we will now discuss, the center-of-mass variables allow us to give a concrete description of a subspace in which the shift-invariant symmetric polynomials must lie. To this effect, it is useful to work with another generating set for the ring of symmetric polynomials,
namely the \emph{power-sum polynomials} in $m$ variables, defined by
\begin{equation}
p_k({\bf y}) = \sum_{i=1}^my_i^k\,.
\end{equation}
Note that every symmetric polynomial can be written as a polynomial in the power-sum polynomials $p_k$. Furthermore, it is easy to see from the definition~\eqref{eq:COMchange} that the first power-sum in `center-of-mass' variables vanishes in such cases,
\begin{equation}\label{eq:COMkernel}
    p_1({\bf y})=0\,.
\end{equation}
As a consequence, we see that, when expressed as a (not necessarily linear) polynomial in power-sum polynomials in center-of-mass variables, a shift-invariant symmetric polynomial does not contain $p_1$.

In particular, it is convenient to choose the power-sum monomials as a basis for the space of shift-invariant polynomials:
\begin{equation}\label{eq:COMpowersums}
    p_{\Delta}({\bf y})=\prod_{i=1}^{l(\Delta)} p_{\Delta_i}({\bf y})\,, \qquad \Delta \in \mathcal{P}^{\hat{1},m}(n)\,.
\end{equation}
The restriction $\Delta_i \leq m$  picks out only algebraically-independent power-sums, just as with usual symmetric functions in $m$ variables. From this argument we see that every shift-invariant symmetric polynomial can be expanded in the basis $\{p_{\Delta}\}$ of power-sum monomials indexed by nuclear partitions $\Delta$. At this point, one still needs to prove that the functions in eq.~\eqref{eq:COMpowersums} are linearly independent. In other words, one needs to prove that the kernel of the map in eq.~\eqref{eq:COMchange} is generated by $p_1$. This statement has been proven in two ways in refs.~\cite{liptrap2008translation,simon2007pseudopotentials}.

In the following, we will restrict the number of variables $m$ to be greater than the size of the partitions, i.e., we will assume $m \geq |\lambda|$.
Hence, the upper bound $\Delta_i \leq m$ on the row length is automatically satisfied, and we are left with $\mathcal{P}^{\hat{1}}(n)$.
Note that eq.~\eqref{eq:COMpowersums} already provides a basis in the space of shift-invariant symmetric polynomials, so in principle it is enough to pass back to functions in $(z,\zb)$ by expressing the functions in eq.~\eqref{eq:COMpowersums} in terms of Schur polynomials, then taking the expectation value in the fishnet matrix model. 
However, as functions of the $\bfx$ variables, the symmetric polynomials in eq.~\eqref{eq:COMpowersums} are quite complicated. In particular, the expansion of this basis in terms of Schur polynomials $s_{\lambda}(\bfx)$ is not known. Nevertheless, this argument demonstrates that it suffices to construct $p^{\hat{1}}(n)$ linearly-independent shift-invariant symmetric polynomials in each degree $n$ and for a fixed value of $m$. 
In the remainder of this section we will present such a set.

\subsection{From shift-invariance to antipodal self-duality}
\label{sec:with_examples}

In this section we will present our main result, namely we will present a formula for a set of $p^{\hat{1}}(n)$ linearly-independent shift-invariant combinations of Schur polynomials. The previous discussion then guarantees that this set forms a basis, and we can then easily obtain a basis for all ASD linear combinations of generalized fishnet integrals. We start by stating our result and we discuss some of its properties. The proof will be presented below.

We denote by $\nu \cup (1^k)$ the partition obtained by concatenating a partition $\nu$ of length $l$ with a string of $1$'s of length $k$:
\begin{equation}
    \nu \cup (1^k) = [\nu_1,\nu_2, \ldots ,\nu_{l},\underbrace{1,\dots ,1}_{k}]\,.
\end{equation}
Then for each nuclear partition $\lambda \in \mathcal{P}^{\hat{1}}(n)$ we define:
\begin{equation}\label{eq:QBasis1}
    W_{\lambda}(\bfx)= \sum_{\substack{\nu \subseteq\lambda\\l(\nu)=l(\lambda)}} \dfrac{(-1)^{|\lambda|-|\nu|} f^{\lambda/\nu}}{\sfC_{{\nu}\cup (1^{|\lambda|-|\nu|})}(m)} \, s_{{\nu}\cup (1^{|\lambda|-|\nu|})}(\bfx)\, ,
\end{equation}
where the quantities appearing in the coefficients have been defined in eqs.~\eqref{eq:flmdet} and~\eqref{eq:ContentProduct}.

Below we prove that the polynomials in eq.~\eqref{eq:QBasis1} are shift-invariant. Let us focus here on proving their linear independence. In fact, they form a basis for the space of all symmetric shift-invariant polynomials. To see that this is true, we note that the sum in eq.~\eqref{eq:QBasis1} runs over all partitions $\nu$ that fit into $\lambda$, with the restriction that the length of $\nu$ is equal to that of $\lambda$. The Schur polynomials that appear in the sum are all of the same degree $|\lambda|$, as exactly $|\lambda|-|\nu|$ extra $1$'s are added to $\nu$. Most importantly, there is a natural one-to-one map between the  polynomials $W_{\lambda}$ and nuclear partitions, because $s_{\lambda}(\bfx)$ is the {\it only} Schur function in the sum~\eqref{eq:QBasis1} that corresponds to a partition from $\mathcal{P}^{\hat{1}}(n)$. All other partitions $\nu$ require appending at least one single-box row, so they are not nuclear. That is, the sum can be written as:
\begin{equation}
    W_{\lambda}(\bfx)=\dfrac{s_{\lambda}(\bfx)}{\sfC_{\lambda}(m)} + \ldots\,,
\end{equation}
where the dots indicate terms that only contain Schur polynomials $s_{\mu}$ with $\mu\notin \cP^{\hat{1}}(n)$. 
Since Schur polynomials form a basis, we conclude that the  polynomials $W_{\lambda}$ are linearly independent. From the previous discussion we know that the dimension of the space of symmetric shift-invariant polynomials attached to partitions from $\cP^{\hat{1}}(n)$ is $p^{\hat{1}}(n)$. Hence, we conclude that the functions $W_{\lambda}$ form a basis for the space of all symmetric shift-invariant polynomials.

We now take expectation values in the fishnet matrix model to define the functions:
\begin{equation}\label{eq:Q_def}
    Q_{\lambda}^{(m)}= \ev{ W_{\lambda}(\bfx) }_{\!\rm F}^{\!(m)}= \sum_{\substack{\nu \subseteq\lambda \\ l(\nu)=l(\lambda)}} \dfrac{(-1)^{|\lambda|-|\nu|} {\color{crimson}f^{\lambda/\nu}}}{\color{azure(colorwheel)}\sfC_{{\nu}\cup (1^{|\lambda|-|\nu|})}(m) } {\color{green(html/cssgreen)}N^{(m)}_{{\nu}\cup (1^{|\lambda|-|\nu|})}} \,,
\end{equation}
where $\lambda\in \cP^{\hat{1}}(n)$ is a nuclear partition and $m\ge |\lambda|$ is an integer.
The color-coding is introduced so that we 
can illustrate the different ingredients with the examples below. Since the $W_{\lambda}(\bfx) $ form a basis for the space of symmetric shift-invariant polynomials, the functions $Q_{\lambda}^{(m)}$ form a basis for all ASD combinations of generalized fishnet integrals. 

\paragraph{Example 1.} For $n=2$, there is only one nuclear partition $\lambda=[2]$. There are two terms in the sum in eq.~\eqref{eq:Q_def}, namely $\nu=[2]$ and $\nu=[1]$. We then obtain:
\begin{equation}
    Q_{[2]}^{(m)} = \dfrac{{\color{green(html/cssgreen)}N_{[2]}^{(m)}}}{{\color{azure(colorwheel)}m(m+1)}} - \dfrac{\color{green(html/cssgreen)}N_{[1,1]}^{(m)}}{{\color{azure(colorwheel)}m(m-1)}}\,.
\end{equation}

 \paragraph{Example 2.}
 For $n=4$ there are two nuclear partitions, $\lambda=[4]$ and $\lambda=[2,2]$. For $\lambda=[4]$ the allowed partitions $\nu$ are $[4],[3],[2],[1]$. Hence:
 \begin{equation}
    \begin{split}
    \label{eq:Q[4](m)}
         Q_{[4]}^{(m)} =&\ \dfrac{ \color{green(html/cssgreen)}N_{[4]}^{(m)}}{\color{azure(colorwheel)}m(m+1)(m+2)(m+3)}- \dfrac{\color{green(html/cssgreen)} N_{[3,1]}^{(m)}}{\color{azure(colorwheel)}m(m+1)(m+2)(m-1)} + \\
         &+\dfrac{\color{green(html/cssgreen)} N_{[2,1,1]}^{(m)}}{\color{azure(colorwheel)}m(m+1)(m-1)(m-2)}- \dfrac{\color{green(html/cssgreen)} N_{[1,1,1,1]}^{(m)}}{\color{azure(colorwheel)}m(m-1)(m-2)(m-3)}\,.
    \end{split}
 \end{equation}
For $[2,2]$ the partitions that appear in the sum are $[2,2]$, $[2,1]$, $[1,1]$:
 \begin{equation}
    \begin{split}
         Q_{[2,2]}^{(m)} =&\ \dfrac{\color{green(html/cssgreen)} N_{[2,2]}^{(m)}}{\color{azure(colorwheel)}m^2(m+1)(m-1)}-\dfrac{\color{green(html/cssgreen)} N_{[2,1,1]}^{(m)}}{\color{azure(colorwheel)}m(m+1)(m-1)(m-2)} + \\
         &+ \dfrac{\color{green(html/cssgreen)} N_{[1,1,1,1]}^{(m)}}{\color{azure(colorwheel)}m(m-1)(m-2)(m-3)}\,.
    \end{split}
 \end{equation}

\begin{figure}[H]
\begin{tikzpicture}[
    every node/.style={inner sep=5pt}
]

{\ytableausetup{boxsize=1.5em}
\scriptsize


\node (a1) at (-0.5,0)
{
$
\begin{ytableau}
\color{blue}0 & \color{blue}1 & \color{blue}2 \\
\color{blue}{-1} & \color{blue}0
\end{ytableau}
$
};

\node (a2) at (2,0)
{
$
\begin{ytableau}
\color{blue}0 & \color{blue}1 & \color{blue}2 \\
\color{blue}{-1} \\
\color{blue}{-2}
\end{ytableau}
$
};

\node (a3) at (4,0)
{
$
\begin{ytableau}
\color{blue}0 & \color{blue}1 \\
\color{blue}{-1} & \color{blue}0 \\
\color{blue}{-2}
\end{ytableau}
$
};

\node (a4) at (6,0)
{
$
\begin{ytableau}
\color{blue}0 & \color{blue}1 \\
\color{blue}{-1} \\
\color{blue}{-2} \\
\color{blue}{-3}
\end{ytableau}
$
};

\node (a5) at (8,0)
{
$
\begin{ytableau}
\color{blue}0 \\
\color{blue}{-1} \\
\color{blue}{-2} \\
\color{blue}{-3} \\
\color{blue}{-4}
\end{ytableau}
$
};


\node (b1) at (-0.5,-3) {$\ydiagram{3,2}$};
\node (b2) at (2,-3) {$\ydiagram{3,1}$};
\node (b3) at (4,-3) {$\ydiagram{2,2}$};
\node (b4) at (6,-3) {$\ydiagram{2,1}$};
\node (b5) at (8,-3) {$\ydiagram{1,1}$};
}

\node (a0) at (-3,0){$\nu \cup (1,\ldots ,1)$};
\node (b0) at (-4,-3) {$\nu \subseteq[3,2]: l(\nu)=2$};
\node[crimson] at (b1.south) [below=6pt,left=2em] {$f^{[3,2]/\nu}$};
\node[crimson] at (b1.south) [below=6pt] {$1$};
\node[crimson] at (b1.south) [below=6pt] {$1$};
\node[crimson] at (b2.south) [below=6pt] {$1$};
\node[crimson] at (b3.south) [below=6pt] {$1$};
\node[crimson] at (b4.south) [below=6pt] {$2$};
\node[crimson] at (b5.south) [below=6pt] {$2$};


\draw[thick] (a1) -- (a2);
\draw[thick] (a2) -- (a3);
\draw[thick] (a3) -- (a4);
\draw[thick] (a4) -- (a5);


\draw[thick] (b1) -- (b2);
\draw[thick] (b2) -- (b3);
\draw[thick] (b3) -- (b4);
\draw[thick] (b4) -- (b5);


\draw[->, thick, shorten >=10pt, shorten <=10pt] (a1) -- (b1);
\draw[->, thick, shorten >=10pt, shorten <=10pt] (a2) -- (b2);
\draw[->, thick, shorten >=10pt, shorten <=10pt] (a3) -- (b3);
\draw[->, thick, shorten >=10pt, shorten <=10pt] (a4) -- (b4);
\draw[->, thick, shorten >=10pt, shorten <=10pt] (a5) -- (b5);

\end{tikzpicture}
\caption{\label{fig:content}The Young diagrams relevant for Example~3 together with the content $j-i$ for each box $(i,j)$.}
\end{figure}

\paragraph{Example 3.} The first example where the path factor is greater than one corresponds to the nuclear partition $[3,2]$. Namely:
 \begin{equation}\label{eq:32example}
    \begin{split}
         Q_{[3,2]}^{(m)}=&\ \dfrac{\color{green(html/cssgreen)} N_{[3,2]}^{(m)}}{\color{azure(colorwheel)}m^2(m+1)(m+2)(m-1)}-\dfrac{\color{green(html/cssgreen)} N_{[3,1,1]}^{(m)}}{\color{azure(colorwheel)}m(m+1)(m+2)(m-1)(m-2)}\\
         &-\dfrac{\color{green(html/cssgreen)} N_{[2,2,1]}^{(m)}}{\color{azure(colorwheel)}m^2(m+1)(m-1)(m-2)} 
         \\
         &+ \dfrac{{\color{crimson}2}\color{green(html/cssgreen)} N_{[2,1,1,1]}^{(m)}}{\color{azure(colorwheel)}m(m+1)(m-1)(m-2)(m-3)}- \dfrac{{\color{crimson}2}\color{green(html/cssgreen)} N_{[1,1,1,1,1]}^{(m)}}{\color{azure(colorwheel)}m(m-1)(m-2)(m-3)(m-4)}\,.
    \end{split}
 \end{equation}

The relevant Young diagrams for the ${\nu}\cup (1^{|\lambda|-|\nu|})$ partitions are shown in figure \ref{fig:content} together with the contents $j-i$ for each box. After shifting by $m$, these are exactly the numbers appearing in the denominators of the expression for the invariant. To count the path number factor {\color{crimson}$f^{\lambda/\nu}$} we also list the ``cuts'' down to $\nu$ of the partitions in the second row. For these, the number of paths to the partition $[3,2]$ should be computed. (The corresponding paths for the partitions $[2,1]$ and $[1,1]$ are shown in  figure~\ref{fig:hasse2} in the appendix.)

For completeness, let us also spell out the antipodal transformation for $N_{[3,2]}^{(m)}$:
\begin{align}
\nonumber\hbfS\big(N_{[3,2]}^{(m)}\big)& = \gr{N_{[3,2]}^{(m)}} + \az{(m+2)}L\gr{N_{[2,2]}^{(m)}} + \az{m}L\gr{N_{[3,1]}^{(m)}} 
        \\
        &+ \dfrac{\cri{2}\az{(m+2)m}}{2!}L^2\gr{N_{[2,1]}^{(m)}} + \dfrac{\az{m(m-1)}}{2!}L^2\gr{N_{[3]}^{(m)}}+\\\nonumber &+\dfrac{\cri{3}\az{(m+2)m(m-1)}}{3!}L^3\gr{N_{[2]}^{(m)}} + \dfrac{\cri{2}\az{(m+2)m(m+1)}}{3!}L^3\gr{N_{[1,1]}^{(m)}}    +
        \\
 \nonumber        &+ \dfrac{\cri{5}\az{(m+2)m(m-1)(m+1)}}{4!}L^4\gr{N_{[1]}^{(m)}} + \dfrac{\cri{5}\az{(m+2)m^2(m-1)(m+1)}}{5!}L^5\gr{N_{\varnothing}^{(m)}}\,.
\end{align}

Similar expansions hold for other determinants $N_{\lambda}^{(m)}$ that appear in eq.~\eqref{eq:32example}. Finally, we can see cancellations in the antipodal transformation for $Q_{[3,2]}^{(m)}$. For example, for the coefficient of the $N_{[3]}^{(m)}$ term after applying the antipode this boils down to the following identity:
\begin{equation}
  \begin{split}
     \underbrace{  \dfrac{ \az{m(m-1)}}{\az{m^2(m+1)(m+2)(m-1)}}}_{[3,2]}- \underbrace{\dfrac{ \az{(m-2)(m-1)}}{\az{m(m+1)(m+2)(m-1)(m-2)}}}_{[3,1,1]} = 0\,.
  \end{split} 
\end{equation}

\subsection{Proof of the shift-invariance}
\label{sec:proof_invariance}

In this subsection, we present the proof that the polynomials $W_{\lambda}(\bfx)$ are shift-invariant.
Consider a generic symmetric, shift-invariant polynomial $A(\bfx)$ and its expansion in terms of Schur polynomials:
\begin{equation}
\label{eq:Asum}
    A(\bfx)= \sum_{\mu} 
\alpha_{\mu} s_{\mu}(\bfx)\,.
\end{equation}
The condition of shift invariance can be rewritten as:
\begin{equation}\label{eq:ShiftInvDer}
    \left( \sum_{i=1}^m \dfrac{\partial}{\partial x_i} \right) A(\bfx)=0\,.
\end{equation}
We can now use eq.~\eqref{eq:shift_s} in infinitesimal form, i.e., we assume that $a$ is infinitesimal and expand eq.~\eqref{eq:shift_s} to linear order in $a$. At this order, the only contributing partitions have Young diagrams that differ from the parent partition by exactly one box; the other partitions are suppressed by at least ${\cal O}(a^2)$.  We find:
\begin{equation}\label{eq:split_s_infinitesimal}
      \left( \sum_{i=1}^m \dfrac{\partial}{\partial x_i} \right)  s_{\mu}(\bfx) =\sum_{ \substack{
      \nu\subset \mu  \\
      |\nu|=|\mu|-1}} \dfrac{\sfC_{\mu}(m)} {\sfC_\nu(m)} \, s_{\nu}(\bfx)\,.
\end{equation}

 Next, we define the sets of \emph{addable} and \emph{removable} boxes of a Young diagram attached to a given partition $\lambda$. We denote these sets by
    $\Add{\lambda}$ and $\Rem{\lambda}$, respectively. Here addable (resp. removable) boxes are such that the Young diagram of a partition with a box added (removed) still has an ordered set of rows, i.e.~a non-increasing sequence of components, and therefore still provides a well-defined Young diagram. In terms of box coordinates, the sets of addable and removable boxes can be defined by
    \begin{equation}
    \begin{split}
    \Add{\lambda } &\,=\left\{ (i,j) \in \ \mathbb{Z}_{+}^{2}\, | j \leq \lambda_{i-1} \,,\lambda_i+1=j \right\}\,,\\ 
    \Rem{\lambda } &\,=\left\{ (i,j) \in \ \mathbb{Z}_{+}^{2}\, | j>\lambda_{i+1} \,,\lambda_i=j \right\}\, ,
    \end{split}\end{equation}
where we set $\lambda_0 = \infty$ and $\lambda_i = 0$ for $i > l(\lambda)$.

For each term on the right-hand side of eq.~\eqref{eq:split_s_infinitesimal},
the content ratio $\sfC_\mu(m)/\sfC_\nu(m)$ has only one factor, associated with the removal of a single box.  Hence it can be written as:
\begin{equation}
      \left( \sum_{i=1}^m \dfrac{\partial}{\partial x_i} \right)  s_{\mu}(\bfx) =\sum_{ \Box \in \Rem{\mu}} (m+j_\Box-i_\Box) s_{\mu-\Box}(\bfx)\,.
\end{equation}
We insert this formula into eq.~\eqref{eq:Asum}, shift the summation variable back to $\mu$, and use the linear independence of the Schur polynomials. Then the invariance condition in eq.~\eqref{eq:ShiftInvDer} for the coefficients $\alpha_\mu$ becomes:
\begin{equation}
    \sum_{\Box \in \Add{\mu}} (m+j_\Box-i_\Box) \alpha_{\mu+\Box} = 0 \,,\quad \forall\mu\,.
\end{equation}
The coefficient can be easily removed by redefining:
\begin{equation}
    \alpha_\mu=\dfrac{\alpha'_\mu}{\sfC_{\mu}(m)}\,,
\end{equation}
which leads to an equation for the new coefficients $\alpha'_\mu$:
\begin{equation}\label{eq:alphasystem}
    \sum_{\Box \in \Add{\mu}} \alpha'_{\mu+\Box} = 0 \,,\quad \forall\mu\,.
\end{equation}
Shift-invariant symmetric functions correspond to linearly independent solutions to eq.~\eqref{eq:alphasystem}.

To write down the solutions to eq.~\eqref{eq:alphasystem}, we need to define a combinatorial operation on partitions. For a pair of partitions $\lambda \in \mathcal{P}^{\hat{1}}$ and $\mu \in \mathcal{P}$, we define the \emph{$\lambda$-cut of $\mu$}, denoted by $\operatorname{cut}_\lambda(\mu)$, as follows: If $\mu$ is longer than $\lambda$, then $\cut_{\lambda}(\mu)$ is the partition obtained by deleting all $1$'s from $\mu$ until either no $1$'s are left or we reach the length of $\lambda$. Formally:
    \begin{equation}
\operatorname{cut}_{\lambda}(\mu)
= \mu \setminus (1^{k}),
\qquad
k = \min\{m_1{(\mu)},\max\{0,\, l(\mu)-l(\lambda)\}\}\,,
    \end{equation}
where $m_1(\mu)$ is the number of $1$'s in $\mu$. 

Let us illustrate this on a few examples. If we consider the pair of partitions represented by the Young diagrams
\begin{equation}
    \mu= \ydiagram{3,2,1,1,1} \textrm{~~~and~~~}  \lambda=\ydiagram{5,3} \,,
\end{equation}
then the $\lambda$-cut of $\mu$ is
\begin{equation}
\cut_{\lambda}(\mu)= \ydiagram{3,2} \,. 
\end{equation}
We can also present an example where there are no $1$'s left before we reach the length of $\lambda$. To this effect, we consider the pair of partitions
\begin{equation}
\mu= \ydiagram{2,2,2,1} \textrm{~~~and~~~} \lambda=\ydiagram{4,3} \,,
\end{equation}
and we obtain
\begin{equation}
\cut_{\lambda}(\mu)= \ydiagram{2,2,2}\,.  
\end{equation}

In the remainder of this section, we show that for every nuclear partition $\lambda$, there is a solution to eq.~\eqref{eq:alphasystem} that is given by
\begin{equation}
    \alpha'_{\lambda\mu} = 
    \begin{cases} (-1)^{|\lambda|-|\cut_{\lambda}(\mu)|} \, f^{\lambda/\cut_{\lambda}(\mu)}\ \ \hbox{for $|\mu|=|\lambda|$}, \\ 0\ \  \hbox{otherwise.} 
    \end{cases}
\end{equation}
The restriction to partitions with $|\mu| = |\lambda|$ is because eq.~\eqref{eq:alphasystem} is homogeneous, so it should be solved by Schur polynomials of the same overall degree. Note also that the coefficients $\alpha'_{\lambda\mu}$ vanish whenever $\cut_\lambda(\mu)$ is not a subpartition of $\lambda$, because in that case $f^{\lambda/\cut_{\lambda}(\mu)} = 0$.

Equation~\eqref{eq:alphasystem} is solved for $\mu=\lambda$ because adding any box leads to a vanishing path number, $f^{\lambda/(\lambda+\Box)} = 0$.
Now we consider $\mu\subset\lambda$. We want to find the $\lambda$-cut of $\mu+\Box$. The sum in eq.~\eqref{eq:alphasystem} is over all possible ways to add a box to $\mu$. There are three types of partitions that can be obtained as a result, and we will consider each case separately. We will illustrate them using $\lambda=[3,3]$ and $\mu=[2,2,1,1]$ as examples:
\begin{equation}
     \mu= \ydiagram[*(airforceblue)]{2,2,1,1} \quad \lambda=\ydiagram[*(mintcream)]{3,3} \rightarrow  \cut_{\lambda}(\mu)= \ydiagram[*(airforceblue)]{2,2} 
\end{equation}
\begin{itemize}
    \item The first case corresponds to the situation where the box is appended to the bottom of the Young diagram. That is, the row coordinate of the added box is $i_\Box=l(\mu)+1$. In this case we have:
\begin{equation}
  \cut_{\lambda}\left({\mu \cup (1)} \right) =\cut_\lambda(\mu)\,,
\end{equation}
because we just make the leg longer. In our example this looks as follows:
\begin{equation}
     \mu= \ydiagram{2,2,1,1} \quad \rightarrow  \quad  \mu \cup (1) = \ydiagram{2,2,1,1}*[*(antiquefuchsia)]{2+0,2+0,1+0,1+0,1} \quad \rightarrow \quad  \cut_{\lambda}(\mu \cup (1))=\cut_{\lambda}(\mu)= \ydiagram{2,2}
\end{equation}
\item The second case corresponds to adding a box to the row with coordinate $i_\Box=l(\lambda)+1$, that is  $(i_\Box,j_\Box)=(l(\lambda)+1,2) \in \Add{\mu}$. This can happen because the last row of the nuclear partition $\lambda$ has length greater than $1$. The box is added the row of $\mu$ that is right below the last row of $\lambda$. In our example this looks like:
\begin{equation}
     \mu= \ydiagram{2,2,1,1} \quad \rightarrow \quad  \mu+\BoxF  = \ydiagram{2,2,1,1}*[*(antiquefuchsia)]{2+0,2+0,1+1} \quad \rightarrow \quad \cut_{\lambda}(\mu +\BoxF )= \ydiagram{2,2,1}*[*(antiquefuchsia)]{2+0,2+0,1+1}
\end{equation}
In this case, we have
\begin{equation}
    \cut_{\lambda}{(\mu+\Box)} =\cut_{\lambda}(\mu) \cup (2)\,,
\end{equation}
i.e., the respective term vanishes, because  $f^{\lambda/ \cut_{\lambda}{(\mu+\Box)} } = 0$ since $\left(\cut_{\lambda}(\mu) \cup (2) \right) \not \subseteq \lambda$.
\item The remaining cases correspond to $i_\Box \leq l(\lambda)$. This means that the box is added to the rows of $\mu$ that lie within $\lambda$. Then we have:
\begin{equation}
    \cut_{\lambda}(\mu+\Box) = \cut_{\lambda}(\mu)+\Box\,.
\end{equation}
In our example only one such box can be added:
\begin{equation}
      \mu= \ydiagram{2,2,1,1} \quad \rightarrow \quad  \mu +{\BoxF  } = \ydiagram{2,2,1,1}*[*(antiquefuchsia)]{2+1,2+0} \quad \Rightarrow \quad \cut_{\lambda}(\mu +\BoxF )=\cut_{\lambda}(\mu)+\BoxF = \ydiagram{2,2}*[*(antiquefuchsia)]{2+1,2+0}
\end{equation}
\end{itemize}

Collecting these three cases and taking into account the sign factor, we get:
\begin{equation}
    f^{\lambda/\cut_{\lambda}(\mu)} - \sum_{i=1}^{l(\lambda)} f^{\lambda/(\cut_{\lambda}(\mu)+\Box_i)} = 0\,,
    \label{eq:path_eqn}
\end{equation}
for $\mu \subset \lambda$,
where by $\Box_i$ we denote the box with coordinate $i_\Box=i$, solving eq.~\eqref{eq:alphasystem}. This relation is nothing but the defining relation for the number of skew-shapes or paths in the Young's lattice (see appendix \ref{sec:appendix}), with the condition that $l(\cut_{\lambda}(\mu))=l(\lambda)$ taken into account. Namely, the terms in the sum represent all paths from $\cut_{\lambda}(\mu)$ to $\lambda$ that pass through $\cut_{\lambda}(\mu)+\Box_i$. But {\it all} paths from $\cut_{\lambda}(\mu)$ to $\lambda$ pass through partitions with an extra box, so eq.~\eqref{eq:path_eqn} holds.
In our example above, there is only one non-vanishing term in the second sum, namely $f^{[3,3]/[3,2]}$. Therefore we have:
\begin{equation}
    f^{[3,3]/[2,2]}-f^{[3,3]/[3,2]}=0\,,
\end{equation}
which holds because $f^{[3,3]/[2,2]}=f^{[3,3]/[3,2]}=1$.

Finally, putting everything together, we obtain the following shift-invariant solutions:
\begin{equation}
    W_\lambda(\bfx) = \sum_{\substack{|\mu| = |\lambda|\\ \cut_{\lambda}(\mu)\subseteq\lambda}
    }   \dfrac{(-1)^{|\lambda|-|\cut_{\lambda}(\mu)|}f^{\lambda/\cut_{\lambda}(\mu)}}{\sfC_{\mu}(m)}s_\mu(\bfx)\,.
\end{equation}
This expression is equivalent to eq.~\eqref{eq:QBasis1} after identifying $\nu = \cut_{\lambda}(\mu)$ and hence $\nu \cup (1^{|\lambda|-|\nu|})=\mu$, which finishes the proof.


\subsection{Cases where $m < |\lambda|$}
\label{sec:m_too_small}

Let us conclude by discussing what happens for $m < |\lambda|$. Indeed, our result for $Q_\lambda^{(m)}$ in eq.~\eqref{eq:Q_def} is only valid for $m\ge |\lambda|$, because the coefficients have explicit singularities for $m < |\lambda|$. This can be very clearly seen from the examples in section \ref{sec:with_examples}. For example, we can see that for $Q_{[\scriptstyle 4]}^{(m)}$ in eq.~\eqref{eq:Q[4](m)} we have $|\lambda|=4$, and the coefficients have poles for $m<4$. 

We find that we can take a linear combination of $Q_{\lambda}^{(m)}$ such that the singular coefficients vanish. 
The general procedure to do this seems to be rather complicated. We thus only illustrate this on an example with $n=|\lambda|=4$ and $m=3$. There are two nuclear partitions, and thus also two invariants, for $n=4$, namely $Q_{[\scriptstyle 4]}^{(m)}$ and $Q_{[\scriptstyle 2,2]}^{(m)}$. We observe that the coefficients of $N_{[1,1,1,1]}^{(m)}$ in both invariants are in fact equal up to a sign. Therefore, taking the sum, we find:
\begin{equation}\label{eq:n4finite}
Q_{[2,2]}^{(m)}+Q_{[4]}^{(m)}=\dfrac{ \color{green(html/cssgreen)}N_{[4]}^{(m)}}{\color{azure(colorwheel)}m(m+1)(m+2)(m+3)}- \dfrac{\color{green(html/cssgreen)} N_{[3,1]}^{(m)}}{\color{azure(colorwheel)}m(m+1)(m+2)(m-1)} +\dfrac{\color{green(html/cssgreen)} N_{[2,2]}^{(m)}}{\color{azure(colorwheel)}m^2(m+1)(m-1)}\,.
\end{equation}
This expression is finite for $m=3$. One can check that there are no other possible finite combinations. Hence at $n=4$ and $m=3$ there is only one antipodal invariant. From the symmetric functions point of view this is expected, because shift-invariant symmetric functions are labeled by nuclear partitions whose components are at most $m$. In our case there is only one such partition $[2,2]$. Hence we could define the ASD function for $m\ge 2$:
\begin{equation}
G_{[2,2]}^{(m)}=Q_{[2,2]}^{(m)}+Q_{[4]}^{(m)}\,.
\end{equation}
Note that by canceling the $N_{[1,1,1,1]}^{(m)}$ term, we also canceled $N_{[2,1,1]}^{(m)}$, which is a priori not a given. Because of this $G_{[2,2]}^{(m)}$ is also finite at $m=2$, as evident from eq.~\eqref{eq:n4finite}.

The extra cancellation can be explained as follows. The fact that the $m$-independent part of the coefficients for $[2,1,1]$ and $[1,1,1,1]$ are equal up to sign follows from eq.~\eqref{eq:alphasystem}. Indeed, let us take $\mu=[\underbrace{1,\ldots, 1}_{n-1}]$. Then the equation looks like:
\begin{equation}
    \alpha'_{[\underbrace{\scriptstyle 1,\ldots, 1}_{n}]}+  \alpha'_{[2,\underbrace{\scriptstyle 1,\ldots, 1}_{n-2}]}=0\,.
\end{equation}
Hence it is the shift-invariance that imposes this coincidence between coefficients.

\section{Conclusions}
\label{sec:Conclusions}

In this paper we constructed an explicit combinatorial formula for many new ASD polynomials in ladder integrals. Each such polynomial is labeled by a nuclear partition, that is an integer partition with all components being greater than one.  The polynomials are linear combinations of certain determinants of ladders, which were originally defined by Coronado in the context of the perturbative expansion of the octagon observable~\cite{Coronado:2018cxj}. Conjecturally, our invariants exhaust all ASD combinations of ladder integrals that also satisfy the Steinmann conditions. 

The key observation that enables our construction is a matrix-model representation of the rectangular fishnet. The existence of such a representation is deeply rooted in integrability properties of fishnet integrals. It maps the antipodal duality to properties of symmetric functions under a common shift of all  variables. Hence, another result of this paper is an explicit description of a new basis in the space of shift-invariant symmetric functions.  A trivial byproduct of our construction is an alternate proof of ASD for square fishnet integrals, which map to the empty partition.

A limitation of our construction is a restriction on the size of the determinants relative to the size of the respective nuclear partitions. It is mapped to a restriction on the number of variables in the symmetric functions. Namely, we dealt with the case $m \geq|\lambda|$. We showed in examples that this restriction can be lifted by considering appropriate linear combinations of the polynomials $Q^{(m)}_{\lambda}$. However, the full description is beyond the scope of this paper and is a matter for future work.

It would be interesting to see if the techniques  used here are applicable in other cases. The most natural set-up are fishnet integrals in two dimensions, which have a similar structure as in  four dimensions. One could also study the octagon from ref.~\cite{Coronado:2018cxj}, which naturally contains sums of ladder determinants, and ask whether the ASD properties are reflected in some form in its perturbative expansion. Finally, it would be very interesting to identify Feynman integrals, or even full observables, that can be expressed in terms of our ASD functions, as this may pave the way to identifying more ASD quantities in QFT.

\vskip0.5cm 

\section*{Acknowledgments}

We thank Benjamin Basso, Francis Brown, Daniel Bump, Max Haensch, Grisha Korchemsky, Didina Serban and Nikita Tselousov for useful conversations.  We thank Saclay IPhT for hosting the meeting {\it Observables in gauge theory and gravity} which stimulated many of the developments in this paper. L.D.~is also grateful for the hospitality of the Max Planck Institute for Physics in Munich.
This work was supported in part by the U.S. Department of Energy (DOE) under Awards No.~DE-FOA-0002705 [KA/OR55/22 (AIHEP)] and~DE-AC02-76SF00515, by the European Union (ERC Consolidator Grant LoCoMotive
101043686) and by the Deutsche Forschungsgemeinschaft (DFG, German Research Foundation) under Germany’s Excellence Strategy – EXC 3107 – Project-ID 53376636. Views and opinions expressed are however those of the
author(s) only and do not necessarily reflect those of the European Union or the European Research Council. Neither the European Union nor the granting authority can be held responsible for them.  The work of F.F.~was supported by a Stanford Major Grant and a Rhodes scholarship.  The research of V.M.~is supported by the Priority 2030 Academic Leadership Initiative, contributing to the educational
work of ``Universities for a New Generation of Leaders'', a project within
the framework of the federal Youth and Children program


\appendix

\section{Symmetric functions and partitions}\label{sec:appendix}

\paragraph{Partitions.}
In this appendix we collect some standard results about symmetric functions and partitions.  More information can be found in e.g.~refs.~\cite{Macdonald:book,stanley2023enumerative2,website}. 
For a partition $\lambda=[\lambda_1, \lambda_2 ,\ldots \ ]$ we call the number of non-zero elements of the partition its length and denote it by $l(\lambda)$. We also denote the sum of the components by $|\lambda|\,=\sum\limits_{i=1}^{l(\lambda)} \lambda_i$. We often refer to partitions $\lambda$ equivalently as Young diagrams. 
One can also define the coordinates $(i,j)$ of a box in a Young diagram with $i$ being the row number and $j$ the column number.  For example, the partition $\lambda=[5,4,2,2]$, depicted below,  has $l(\lambda)=4$ and $|\lambda|=13$:
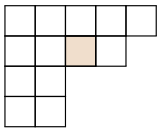
\vspace{0.3cm}
\begin{center}
\ytableausetup{boxsize=1em,aligntableaux = center}  \quad \ydiagram{5,4,2,2}*[*(almond) ]{0,2+1}
\end{center}
\vspace{0.3cm}
The shaded box in the diagram above has coordinates $(2,3)$.  The difference $j-i$ is called the \emph{content} of a given box of a Young diagram. The content of the highlighted box is equal to $3-2=1$. 

Other important objects are skew-shaped partitions. We say that $\mu$ is contained in $\lambda$, $\mu \subseteq\lambda$ if $\mu_i \leq \lambda_i$ for all $i$. Then the skew diagram $\lambda/\mu$ is the set of boxes of $\lambda$ that are not boxes of $\mu$. 

\paragraph{Young's lattice and paths.}
The inclusion $\mu \subseteq \lambda$ induces a partial order on partitions. The resulting poset of partitions is called a Young lattice. The construction of the ASD invariants and the transformation rules can be visualized using the Hasse diagram of the Young lattice. It is a directed graph, where nodes are labeled by partitions. Two nodes are connected if $\mu\subset \lambda$ and $|\lambda|=|\mu|+1$, and the direction of the edges is chosen to be from the smaller partition to the larger one. This means that an edge is directed from a partition $\mu$ to a partition that is obtained by adding a single box to its Young diagram. The Hasse diagram up to degree 4 is shown in figure~\ref{fig:HasseX}.

\begin{figure}[!h]
\begin{center}
    \begin{tikzpicture}[
    every node/.style={inner sep=5pt},
    level distance=1.8cm,
    sibling distance=2.2cm
]


\node (e) at (0,-0.5) {$\varnothing$};

\node (1) at (0,-1.5) {$\ydiagram{1}$};

\node (2) at (-1,-3) {$\ydiagram{2}$};
\node (11) at (1,-3) {$\ydiagram{1,1}$};

\node (3) at (-3,-4.5) {$\ydiagram{3}$};
\node (21) at (0,-4.5) {$\ydiagram{2,1}$};
\node (111) at (3,-4.5) {$\ydiagram{1,1,1}$};

\node (4) at (-4,-6.5) {$\ydiagram{4}$};
\node (31) at (-2,-6.5) {$\ydiagram{3,1}$};
\node (22) at (0,-6.5) {$\ydiagram{2,2}$};
\node (211) at (2,-6.5) {$\ydiagram{2,1,1}$};
\node (1111) at (4,-6.5) {$\ydiagram{1,1,1,1}$};


\draw (e) -- (1);

\draw (1) -- (2);
\draw (1) -- (11);

\draw (2) -- (3);
\draw (2) -- (21);

\draw (11) -- (21);
\draw (11) -- (111);

\draw (3) -- (4);
\draw (3) -- (31);

\draw (21) -- (31);
\draw (21) -- (22);
\draw (21) -- (211);

\draw (111) -- (211);
\draw (111) -- (1111);
\end{tikzpicture}

\end{center}
\caption{\label{fig:HasseX}
The Hasse diagram of the Young's lattice up to degree four.}
\end{figure}

The number of downward paths in the Young's lattice that start at $\mu$ and end at $\lambda$ can be computed explicitly by Aitken's determinantal formula (Corollary 7.16.3 in ref.~\cite{stanley2023enumerative2}):
\begin{equation}\label{eq:flmdetapp}
        f^{\lambda/\mu} = (|\lambda|-|\mu|)!  \,\det \left( \dfrac{1}{(\lambda_i-\mu_j+j-i)!} \right) \,,
    \end{equation}
where $1/k!$ is defined to be $0$ if $k<0$ for $k=\lambda_i-\mu_j+j-i$.

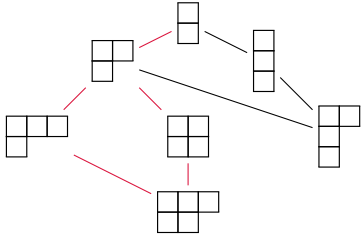
\begin{figure}[hh]
\begin{center}
\begin{tikzpicture}[
    every node/.style={inner sep=2pt}
]

\ytableausetup{boxsize=0.66em}
\node (11a) at (0,0) {$\ydiagram{1,1}$};

\node (21a) at (-1,-0.5) {$\ydiagram{2,1}$};
\node (111a) at (1,-0.5) {$\ydiagram{1,1,1}$};

\node (31a) at (-2,-1.5) {$\ydiagram{3,1}$};
\node (22a) at (0,-1.5) {$\ydiagram{2,2}$};
\node (211a) at (2,-1.5) {$\ydiagram{2,1,1}$};

\node (32a) at (0,-2.5) {$\ydiagram{3,2}$};


\draw[crimson] (11a) -- (21a);
\draw (11a) -- (111a);

\draw[crimson] (21a) -- (31a);
\draw[crimson] (21a) -- (22a);
\draw (21a) -- (211a);

\draw (111a) -- (211a);

\draw[crimson] (31a) -- (32a);
\draw[crimson] (22a) -- (32a);

\end{tikzpicture}
\end{center}
    \caption{A section of the Hasse diagram of the Young lattice. Red denotes the paths between partitions [1,1] and [3,2].}
    \label{fig:hasse2}
\end{figure}

As an example, let us look at a portion of the lattice corresponding to the paths between the partitions $[1,1]$ and $[3,2]$ in figure~\ref{fig:hasse2}. Clearly, there are two paths, giving $f^{[3,2]/[1,1]}=2$ in agreement with the general formula.  This factor accounts for the factor of {\color{crimson}2} in the last term in eq.~\eqref{eq:32example}.

\paragraph{Symmetric functions and Schur polynomials}

The ring of symmetric polynomials in $m$ variables $\bfx=(x_1,\ldots, x_m)$ is graded by the total degree. There are several canonical bases often used in the theory of symmetric functions \cite{Macdonald:book}. Any  basis in the ring of symmetric polynomials is labelled by partitions. The ones that are relevant for us are the power-sum and the Schur polynomial bases. Power-sums are defined by
\begin{equation}
    p_{k}(\bfx)=\sum_{i=1}^m x_i^k\,.
\end{equation}
Power sums with $k \leq m$ are algebraically independent and generate (multiplicatively) the ring of symmetric polynomials in $m$ variables. To find a linear basis one takes generic monomials of power-sums, which can be labeled by partitions $\Delta = [\Delta_1 ,\Delta_2,\ldots]$:
\begin{equation}
    p_\Delta(\bfx) = \prod_{i=1}^{l(\Delta)} p_{\Delta_{i}}(\bfx)\,.
\end{equation}
At fixed $m$ power-sums with $k > m$ are algebraically related to those with $k \leq m$. 

The other basis that we use is the Schur-polynomial basis. Schur polynomials are defined by formula \eqref{eq:Schur} and depend on the number of variables $m$ through the size of the determinant. It is useful to write these polynomials in the power sum basis, where the $m$ dependence is implicit. A few examples of Schur polynomials at low degree are
\begin{equation}
    s_{[2]} = \dfrac{p_2(\bfx)+p_1(\bfx)^2}{2} \,, \quad s_{[1,1]} = \dfrac{p_1^2(\bfx)-p_2(\bfx)}{2} \,.
\end{equation}
Schur polynomials are homogeneous, with the degree of $s_{\lambda}(\bfx)$ being $|\lambda|$, the total number of boxes in $\lambda$. One can prove that Schur polynomials also form a basis in the space of symmetric polynomials. The constraint on the degree of the power-sum now translates into the constraint on the length of $\lambda$. In particular one has:
\begin{equation}
    s_{\lambda}(\bfx)=0 \quad\Leftrightarrow  \quad l(\lambda) > m\,.
\end{equation}
At the same time one can show that all non-vanishing Schur polynomials are linearly independent and hence form a basis. One can verify that, by transposing the Young diagram along the diagonal, the two constraints on partitions (on the size of their components, from the power-sum basis, and on their total length, from the Schur-polynomial basis) give the same number of elements.

\providecommand{\href}[2]{#2}\begingroup\raggedright\endgroup


\end{document}